\documentclass[a4paper,11pt]{article}
\usepackage{jcappub} 
\usepackage{lineno}
\usepackage{bm}

\usepackage{booktabs}

\title{Cosmological Model Independent Constraints on Lorentz Invariance Violation with Updated Gamma-Ray Burst Observations: An Artificial Neural Network Approach}

\author[a,b]{Jun Tian}
\author[a,1]{Yu Pan\note{Corresponding author.}}
\author[c,d,2]{Shuo Cao\note{Corresponding author.}}
\author[b]{Qing-Quan Jiang}
\author[e,f]{Wei-Liang Qian}

\affiliation[a]{School of Electronic Science and Engineering, Chongqing University of Posts and Telecommunications, Chongqing 400065, China}
\affiliation[b]{ School of Physics and Astronomy, China West Normal University, Nanchong 637002, China}
\affiliation[c]{ Institute for Frontiers in Astronomy and Astrophysics, Beijing Normal University, Beijing 102206, China}
\affiliation[d]{School of Physics and Astronomy, Beijing Normal University, Beijing 100875, China}
\affiliation[e]{Escola de Engenharia de Lorena, Universidade de S{\~a}o Paulo 12602-810, Lorena, SP, Brazil}
\affiliation[f]{Center for Gravitation and Cosmology, College of Physical Science and Technology, Yangzhou University, Yangzhou 225009, China}

\emailAdd{panyu@cqupt.deu.cn, caoshuo@bnu.edu.cn}

\abstract{Searching for Lorentz invariance violation (LIV) using astrophysical sources such as gamma-ray bursts (GRBs) is crucial for probing quantum gravity. However, the dependence of LIV constraints on assumed cosmological models has been largely overlooked. In this work, we present a model-independent reconstruction of the cosmic expansion history using artificial neural networks (ANN), thereby avoiding biases from specific cosmological priors. We analyze 74 GRB time delays, including 37 measurements from GRB~160625B across multiple energy bands at $z = 1.41$, and 37 additional bursts spanning redshifts $0.117 \leq z \leq 1.99$. Our analysis yields stringent constraints on both linear and quadratic LIV, with $E_{\mathrm{QG},1} \geq 2.60 \times 10^{15}~\mathrm{GeV}$ and $E_{\mathrm{QG},2} \geq 1.21 \times 10^{10}~\mathrm{GeV}$. The linear limit is within four orders of magnitude of the Planck scale. By leveraging a large sample of GRBs, our approach significantly enhances the robustness of LIV constraints, providing a powerful, cosmological-independent framework for future tests of quantum gravity.}

\begin{document}

\maketitle
\flushbottom

\section{Introduction}
\label{section1}

The Lorentz invariance holds crucial significance in physics and is a fundamental assumption of relativity. However, several quantum gravity (QG) theories suggest that Lorentz invariance violation (LIV) may occur at the Planck energy scale, $E_{\mathrm{QG}} \approx E_{\mathrm{Pl}}  = \sqrt{\hbar c^{5}/G }\simeq  1.22\times 10^{19} \ \mathrm{GeV}$ \cite{Mattingly(2005),Amelino-Camelia(2013)}.
Astrophysical observations of gamma-ray bursts (GRBs) offer a unique opportunity to probe potential LIV effects predicted by QG theories. GRBs are the brightest astrophysical events in the Universe, characterized by extensive distances, brief spectral lags, and energetic emissions \cite{Amelino-Camelia(2013),Ellis(2008),Jacob(2008)}. They stand out as the most promising candidates for conducting LIV tests \cite{Amelino-Camelia(2013)}. Over the past two decades, GRBs have served as the primary source for a majority of LIV tests \cite{Desai(2023),Liao(2022)}.

GRBs are divided into two types based on their duration: short GRBs (SGRBs), lasting less than two seconds, and long GRBs (LGRBs), lasting more than two seconds \cite{Kouveliotou(1993)}.
SGRBs are commonly linked to neutron star mergers \cite{Nakar(2007)}, while LGRBs are typically associated with core-collapse supernovae \cite{Woosley(2006)}.
LGRBs typically exhibit more pronounced positive or negative spectral lags compared to SGRBs \cite{Chen(2005),Ukwatta(2012),Bernardini(2015)}, whereas SGRBs generally display smaller spectral lags \cite{Desai(2023),Bernardini(2015)}. Spectral lag, a well-established phenomenon in GRBs, is defined as the time delay between the arrival of low-energy and high-energy photons in the light curve \cite{Xiao(2022)}. The primary approach for searching LIV with GRBs is analyzing these spectral lags \cite{Desai(2023)}.

The lower limits of $E_{\mathrm{QG}}$ in some LIV tests have approached the Planck energy scale, determined through the observation of spectral lag in extremely high-energy photons from individual GRBs \cite{Abdo(2009)A,Abdo(2009)B,Acciari(2020)}. However, spectral lag includes both the time delay associated with LIV and an unknown intrinsic time delay. Using a single burst to constrain LIV makes it challenging to disentangle them. Ellis et al. \cite{Ellis(2008)} addressed this issue by stacking a sample of GRBs, primarily LGRBs, and replacing the intrinsic time delay in the rest frame with an assumed constant for all GRBs. Nevertheless, it is unlikely that every GRB shares an identical intrinsic spectral lag. Wei et al. \cite{Wei(2017)} employed a more realistic power-law model using multiple lags from GRB 160625B, explaining the transition in spectral lags from positive to negative. However, the data fitting in their study was not entirely satisfactory.
Agrawal et al. \cite{Agrawal(2021)} employed a similar approach, stacking a larger number of GRBs, which improved the precision of the results, but also faced challenges related to the goodness of the data fitting.

Numerous studies have employed cosmological models, such as the $\Lambda$CDM model, to constrain LIV.
However, these models are based on the general relativity framework without considering LIV \cite{Pan(2020)}. Furthermore, the latest release from the Dark Energy Spectroscopic Instrument (DESI) Data Release 2 (DR2) provides evidence for dynamical dark energy, suggesting that modifications to the existing $\Lambda$CDM model might be necessary \cite{2025arXiv250314739D,2025arXiv250314738D,2025arXiv250314743L}.
Additionally, recent findings show the effects of different cosmological models on LIV constraints, revealing the complexity of this relationship and adding more uncertainty to the results \cite{Staicova(2023)}. 
To address this issue, Pan et al. \cite{Pan(2020)} first employed a model-independent method using the Gaussian Process (GP) to reconstruct cosmic expansion history (represented by the Hubble parameter $H(z)$) for constraints on LIV. Their results are more robust and conservative. The GP method has been widely used in cosmology \cite{2018ApJ...856....3Y,2019ApJ...886...94L,2023MNRAS.523.3406F}.
Nevertheless, GP has various disadvantages, such as overfitting and a heavy dependence on the selection of covariance functions \cite{Mukherjee(2022)}. Recent research has also emphasized the need for caution when using GP for the reconstruction of the Hubble parameter \cite{Zhou(2019)}.
In contrast, an artificial neural network (ANN) is a fully data-driven method that accurately captures the input data distribution with a suitable network architecture \cite{Wang(2020)}.
Recent studies have increasingly utilized this method to reconstruct the Hubble parameter \cite{Wang(2020),Go'mez-Vargas(2023),Qi(2023)}.

In this paper, we propose a novel, model-independent method using an ANN to reconstruct the cosmic expansion history, which is subsequently used to constrain LIV. This approach effectively circumvents the potential influence of cosmological models. The theoretical framework for LIV is introduced in Section~\ref{sec:2.1}, the observational datasets are presented in Section~\ref{sec:2.2}, and the ANN method is described in Section~\ref{sec:2.3}. Results and discussion are presented in Section~\ref{section3}. Finally, the main conclusions are summarized in Section~\ref{section4}.

\section{Methodology and Datasets}\label{section2}
\subsection{Lorentz Invariance Violation}\label{sec:2.1}
In some QG models, Lorentz invariance may be violated around the Planck energy scale $E_{\mathrm{QG}} \sim   10^{19} \ \mathrm{GeV}$, leading the speed of light to depend on the energy of photons \cite{Amelino-Camelia(2013),1998Natur.395Q.525A}.
This phenomenon can be described by a modified dispersion relation (MDR). Although MDR does not presuppose any specific theoretical framework, it enables a variety of experimental tests \cite{2008ApJ...689L...1K,2022PrPNP.12503948A}.
In our work, we use an empirical formula for the MDR of massless particles, widely recognized for its robustness across various theoretical contexts \cite{2022PrPNP.12503948A}. This relation can be expressed as follows:
\begin{equation}
E^{2}=c^{2}p^{2}\left [ 1+f(E/E_{\mathrm{QG}}) \right ] ,
\label{eq:1}
\end{equation}
where $E_{\mathrm{QG}}$ denotes the Planck energy scale, $f$ is a model-dependent function of the dimensionless ratio of $E/E_{\mathrm{QG}}$. The dispersion relation can be expanded as a Taylor series \cite{2013PhRvD..87l2001V}:
\begin{equation}
E^{2}\simeq  c^{2}p^{2}\left [ 1 - \sum_{n=1}^{\infty }  s_{\pm }\left ( \frac{E}{E_{\mathrm{QG},n}}  \right )^{n}  \right ] ,
\label{eq:2}
\end{equation}
For $E_{\mathrm{QG},n} \gg E$, the lowest order term in the series is expected to dominate the sum, so we keep only the lowest-order dominant term:
\begin{equation}
E^{2} \simeq  c^{2}p^{2}\left [ 1 - s_{\pm }\left ( \frac{E}{E_{\mathrm{QG},n} }  \right )^{n}    \right ],
\label{eq:3}
\end{equation}
Assuming the relation $v(E)=\partial E /\partial p$ holds at least approximately, the energy-dependent velocity of light is \cite{1998Natur.395Q.525A}:
\begin{equation}
v(E)=c\left [ 1-s_{\pm }\frac{n+1}{2}\left ( \frac{E}{E_{\mathrm{QG}} }    \right )^{n}     \right ].
\label{eq:4}
\end{equation}
Here, $s_{\pm} = +1$ (or $-1$) indicates whether the photon velocity decreases (or increases) with the growth of photon energy, and $n$ is a term dependent on the model, with $n = 1$ (or $2$) corresponding to a linear (or quadratic) term.
Considering that most QG models predict lower-energy photons generally travel faster than higher-energy ones due to LIV effects \cite{1998Natur.395Q.525A,1995AnPhy.243...90L,1997IJMPA..12..607A}, we only focus on $s_{\pm } = +1$ in this work.

Due to the relationship between the velocity of light and energy, photons of varying energies emitted by the identical origin will reach Earth at distinct moments.
Recent astronomical observations, particularly type Ia supernovae (SNe Ia), indicate that the universe contains a negative pressure component that drives its accelerated expansion \cite{1998AJ....116.1009R}. Taking into account the cosmological expansion, the LIV-induced time delay between two photons with observed energies $E$ (high-energy) and $E_0$ (low-energy), emitted from a source at redshift $z$, is given by~\cite{Jacob(2008),Zhang(2015)}:
\begin{equation}
\Delta  t_{\mathrm{LIV}}= -\frac{1+n}{2H_{0} } \frac{E^{n}-E_{0}^{n}}{E_{\mathrm{QG},n}^{n} }  \int_{0}^{z}\frac{(1+z{'})^{n}dz{'}  }{h(z{'} )},
\label{eq:5}
\end{equation}
where $H_{0}$ is the Hubble constant, and $h(z) = H(z)/H_{0}$ denotes the dimensionless Hubble expansion rate at redshift z.

The complete time delay across various energy bands consists of five terms:
\begin{equation}
\Delta t = \Delta t_{\mathrm{LIV}} + \Delta t_{\mathrm{int}} + \Delta t_{\mathrm{spe}} + \Delta t_{\mathrm{DM}} + \Delta t_{\mathrm{gra}},
\label{eq:6}
\end{equation}
\noindent
where $\Delta t_{\mathrm{LIV}}$ represents the potential time delay due to LIV via an energy-dependent velocity of light, $\Delta t_{\mathrm{int}}$ is the intrinsic time delay between two test photons, and this has the largest uncertainty since it is impossible to determine it based on the observational data alone, $\Delta t_{\mathrm{spe}}$ is the potential time delay due to special relativistic effects when the photons have a non-zero rest mass, $\Delta t_{\mathrm{DM}}$ denotes the time delay contribution from the dispersion by the line-of-sight free electron content, and $\Delta t_{\mathrm{gra}}$ is the gravitational potential along the propagation path of photons if the Einstein equivalence principle is violated. However, recent works have pointed out that, for cosmic transient sources (such as GRBs), the last three terms  have no significant influence on the total time delay $\Delta t$, and can be ignored \cite{Gao(2015),Wei(2015)}. Thus, we assume the $\Delta t$ is the sum of $\Delta t_{\mathrm{LIV}}$ and $\Delta t_{\mathrm{int}}$.

Specifically, our current understanding of the detailed emission mechanism of GRBs remains limited. While some astrophysical models, such as the structured jet model, can qualitatively explain the existence of intrinsic time delays, they often fail to accurately quantify or predict their specific behavior \cite{2024ApJ...975L..29V}. This underscores the importance of selecting an appropriate parameterization for intrinsic time delay.
Traditionally, the intrinsic time delay is modeled as $\Delta t_{\mathrm{int}} = b(1+z)$, where $b$ is assumed to be a constant across bursts \cite{Ellis(2008)}. This formulation assumes a uniform intrinsic time delay across all GRBs. However, this assumption may not universally hold.
For SGRBs, using $b$ to represent the intrinsic time delay may be reasonable due to small spectral lags \cite{Xiao(2022)}.
However, LGRBs have longer spectral lags and are very likely to have different intrinsic time lags, the format of a power-law function is more reasonable \cite{Wei(2017)}. 
In the rest frame of the GRB source, we adopt the following form:
\begin{equation}
\Delta t_{\mathrm{int}}^{\mathrm{rest}} = \tau \left [ \left ( \frac{E_{\mathrm{rest}}}{\mathrm{keV}}\right )^{\alpha } -\left ( \frac{E_{\mathrm{rest},0} }{\mathrm{keV}} \right )^{\alpha } \right ],
\label{eq:7}
\end{equation}
where $\tau$ and $\alpha$ are free parameters, and $E_{\mathrm{rest}}$ and $E_{\mathrm{rest},0}$ represent the high- and low-energy photons in the source rest frame, respectively.
Accounting for cosmological redshift $z$, the observed time delay and photon energy are related by $\Delta t_{\mathrm{int}}^{\mathrm{obs}} = \Delta t_{\mathrm{int}}^{\mathrm{rest}} (1+z)$ and $E_{\mathrm{rest}} = E_{\mathrm{obs}} (1+z)$. Substituting into Eq.~\eqref{eq:7}, we obtain:
\begin{equation}
\Delta t_{\mathrm{int}}^{\mathrm{obs}} = \tau \left [ \left ( \frac{E}{\mathrm{keV}} \right )^{\alpha } - \left ( \frac{E_{0}}{\mathrm{keV}} \right )^{\alpha } \right ] (1+z)^{1+\alpha },
\label{eq:8}
\end{equation}

The total theoretical time delay between photons of observed energies $E$ and $E_0$ is obtained by combining the LIV-induced delay from Eq.~\eqref{eq:5} and the intrinsic delay model in Eq.~\eqref{eq:8}:
\begin{equation}
\Delta t_{\mathrm{th}} = \Delta t_{\mathrm{LIV}} + \Delta t_{\mathrm{int}}^{\mathrm{obs}}.
\label{eq:9}
\end{equation}
For convenience in parameter estimation, Eq.~\eqref{eq:9} can be rewritten as:
\begin{equation}
\Delta t_{\mathrm{th}} = -\left( a_{\mathrm{LIV},n} \right)^{n} \cdot \mathcal{K}_n + \Delta t_{\mathrm{int}}^{\mathrm{obs}},
\label{eq:10}
\end{equation}
where $a_{\mathrm{LIV},n} \equiv 1 / E_{\mathrm{QG},n}$ is introduced as a rescaled parameter that is more amenable to observational constraints, since $E_{\mathrm{QG},n}$ is expected to be close to or above the Planck scale.
The kernel $\mathcal{K}_n$ encapsulates the energy dependence and cosmological propagation effect:
\begin{equation}
\mathcal{K}_n = \frac{(1+n)(E^n - E_0^n)}{2H_0} \int_0^z \frac{(1+z')^n}{h(z')}  dz',
\label{eq:11}
\end{equation}
This formulation allows us to constrain $a_{\mathrm{LIV},n}$ directly from the data, where $a_{\mathrm{LIV},n} = 0$ is equivalent to no LIV effect. In the following analysis, we treat $a_{\mathrm{LIV},n}$ as a free parameter to be constrained by the observed time delays across the GRB sample.

Lastly, we emphasize that many previous studies have relied on specific cosmological models, such as flat $\Lambda$CDM model to obtain the function $h(z)$. However, this is cosmological model-dependent. It is shown that the dimensionless Hubble parameter functions $h(z)$ of different cosmological models differ significantly at high redshift, which will significantly affect our study of LIV. The test of LIV by using cosmological model independent method is particularly important.

\subsection{Observational Data}\label{sec:2.2}
In this paper, we have selected a sample of 74 time delay measurements for GRBs (mostly LGRBs), which comprises three datasets analyzed in \cite{Wei(2017),Ellis(2008),Xiao(2022)}.
These time delays were obtained by measuring the time lags between various energy bands within their respective light curves.

The first dataset is presented in Table 1 of Ellis et al. \cite{Ellis(2008)}. This table contains 35 time delays for GRBs at various redshifts, primarily LGRBs, within the redshift range of $0.168 \sim 6.29$.
Among these GRBs, there are 9 GRBs from BATSE data (time resolution: 64 ms), 15 GRBs from HETE data (time resolution: 164 ms), and 11 GRBs from SWIFT data (time resolution: 64 ms).
Spectral lags were calculated by analyzing the light curves, comparing the $115-350 \ \mathrm{keV}$ energy band with the $25-55 \ \mathrm{keV}$ energy band.
However, it should be noted that in our analysis, we have only selected the data with redshifts less than 2.
The reason for this choice will be explained in the next subsection.
Therefore, the first dataset consists of 21 time delays with a redshift range of $0.168 \sim 1.99$.

The second dataset consists of 37 time delays from GRB 160625B at various energy bands (all with a common redshift value of $z = 1.41$), is presented in Table 1 of Wei et al. \cite{Wei(2017)}
GRB 160625B was first identified by the Fermi Gamma-Ray Burst Monitor (GBM) and subsequently confirmed by the Fermi Large Area Telescope (LAT).
GRB 160625B's light curve displays three distinct sub-pulses and the total duration of approximately $T_{90} = 770$s \cite{zhang(2017)}, which means this burst is LGRB.
Specifically, the sub-pulse number two is exceptionally luminous, enabling straightforward extraction of the light curves in various energy bands.
By analyzing the light curves, the spectral lags were calculated using the CCF method in the lowest energy band $10-12 \ \mathrm{keV}$ relative to other higher energy bands $15-350 \ \mathrm{keV}$.

The last dataset is presented in Table 3 of Xiao et al. \cite{Xiao(2022)}.
This table contains data on 21 time delays for LGRBs, with redshifts covering the range from z = 0.117 to z = 2.938 (redshifts obtained from Swift Burst Analyser \cite{Evans(2009)}, Fermi GBM Burst Catalog \cite{Kienlin(2020)}, and GCN circulars of the Swift and GBM teams).
These LGRBs are observed by GBM, BAT, and GECAMB detectors, but only BAT observed all LGRBs, so we chose to utilize BAT data for our analysis.
These spectral lags were extracted using the Li-CCF method, which calculates the spectral lags in light curves recorded in energy bands ranging from $15-70\ \mathrm{keV}$ to $120-250\ \mathrm{keV}$.
As previously mentioned, we limited our selection to data with redshifts less than 2, resulting in a dataset of 16 time delays with redshifts in the range of $0.117 \sim 1.937$.

Finally, the combined dataset consists of 37 time delays from GRB 160625B at various energy bands, all sharing a common redshift value of $z = 1.41$ \cite{Wei(2017)}, as well as 37 time delays from various other GRBs with redshift values ranging from $z = 0.117$ to $z = 1.99$ \cite{Ellis(2008),Xiao(2022)}.

\subsection{Artificial Neural Network  Method}\label{sec:2.3}

In this work, we employ an ANN to reconstruct the cosmic expansion history, represented by the Hubble parameter $H(z)$.
A typical ANN comprises multiple layers of interconnected nodes, including an input layer, one or more hidden layers, and an output layer. With a suitably chosen architecture, ANN has the capacity to approximate any continuous function \cite{HORNIK1989359}. Our implementation is based on \textbf{ReFANN}, a Python-based code developed by Wang et al. \cite{Wang(2020)}, which utilizes a single hidden layer comprising 4096 neurons. The network takes the redshift $z$ as input and outputs the corresponding Hubble parameter $H(z)$ along with its associated uncertainties.
We adopt the Exponential Linear Unit (ELU) as the activation function in the hidden layer. Optimization is carried out using the Adam algorithm with an initial learning rate of 0.003. The loss function is defined as the least absolute deviation. The model is implemented using \textbf{PyTorch}, a widely used open-source framework for tensor computation and deep learning. Training is carried out over 30,000 epochs, and the final model is selected based on the lowest achieved training loss.

The training dataset consists of 32 measurements of the Hubble parameter $H(z)$ obtained from cosmic chronometers.
To ensure complete model independence of the $H(z)$ functions, it is essential to carefully select the data, avoiding any influence from the cosmological model.
The values of $H(z)$ can be acquired using two distinct methods.
The first method is the cosmic chronometer method, which calculates the age differences among galaxies undergoing passive evolution at various redshifts \cite{Jimenez(2002)}.
The Hubble parameter of the cosmic chronometer is denoted as CC $H(z)$, and they offer model-independent measurements of $H(z)$ \cite{Jimenez(2002)}.
Another one is through the detection of radial BAO features \cite{Go'mez-Vargas(2023),Blake(2012),Samushia(2013)}.
However, the values acquired by this method depend on a presupposed fiducial cosmological model.
Given the requirement for model-independent $H(z)$ reconstruction, we choose the CC $H(z)$ data for this purpose.
We utilized the latest collection of 32 data points of CC $H(z)$ \cite{Zhang(2014),Stern(2010),Moresco(2012),Moresco(2016),Ratsimbazafy(2017)}, spanning the redshift range extending to approximately $z = 2$.

\begin{figure}
  \begin{center}
  \includegraphics[width=0.7\hsize]{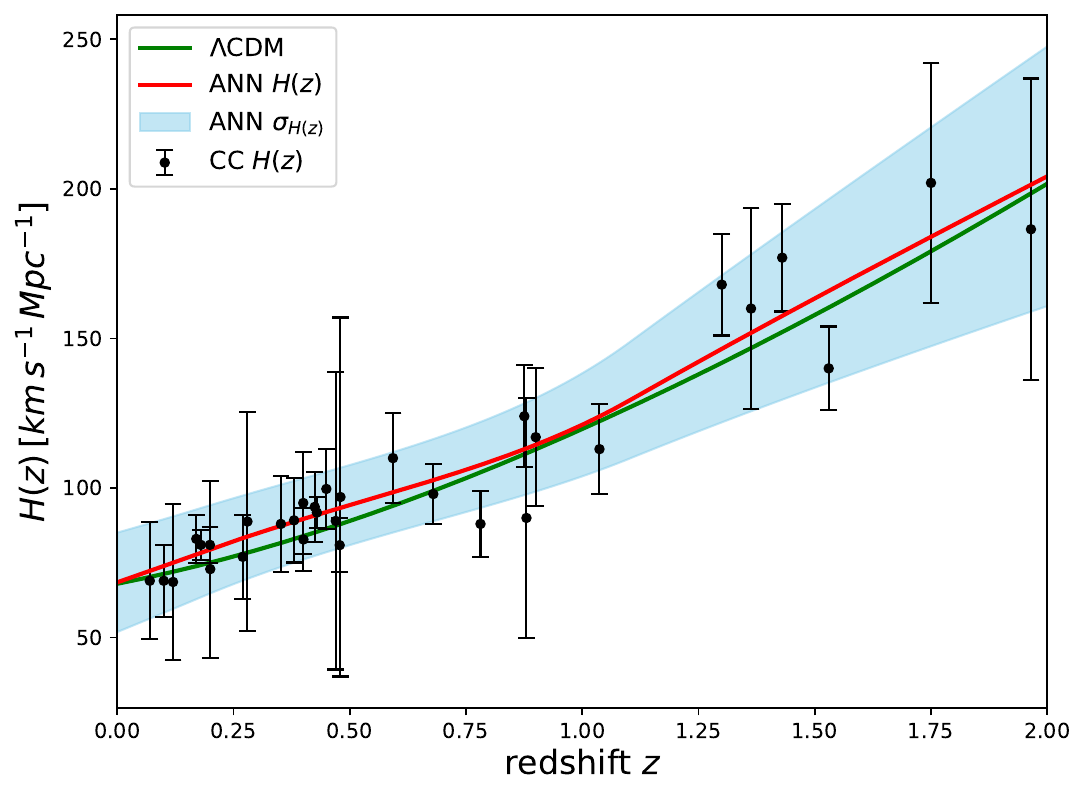}
  \end{center}
  \caption{The reconstructed $H(z)$ (shown as a red curve) using ANN, along with the corresponding $1\sigma$ error (depicted as a sky-blue area), and the CC $H(z)$ data (represented by black dots with error bars). Additionally, the green curve represents the theoretical prediction of the standard cosmological model.  \label{example1}}
  \end{figure}

Figure~\ref{example1} shows the results obtained from the ANN-based reconstruction of the Hubble parameter, denoted as ANN~$H(z)$.
In this figure, black dots with error bars represent CC~$H(z)$ data, and the red curve and sky-blue areas indicate the ANN~$H(z)$'s best values and $1\sigma$ errors. 
For comparison, the green curve shows the prediction of the standard $\Lambda$CDM model, defined as $ H(z) = H_0 \sqrt{\Omega_m (1+z)^3 + 1 - \Omega_m}$,
with $H_0 = 70~\mathrm{km~s^{-1}~Mpc^{-1}}$ and $\Omega_m = 0.3$.
It is important to note that all CC~$H(z)$ data is consistent with the sky-blue areas, taking into account the $1\sigma$ error, which indicates the success of ANN~$H(z)$.
The reconstructed Hubble constant is determined as $H_0 = 68.48 \pm 16.61~\mathrm{km~s^{-1}~Mpc^{-1}}$, a value very close to the Planck CMB result: $H_0 = 67.4~\mathrm{km~s^{-1}~Mpc^{-1}}$~\cite{Aghanim(2021)}.
This similarity further validates the accuracy of the reconstructed Hubble parameter.
Furthermore, the reconstructed ANN~$H(z)$ exhibits a behavior that closely follows the trend predicted by the $\Lambda$CDM model. Importantly, the ANN~$H(z)$ reconstruction is entirely data-driven and does not rely on any underlying physical assumptions or cosmological priors.

Based on this result, the dimensionless Hubble expansion rate $h(z) =$ ANN $H(z) / H_{0}$.
However, it is important to note that the CC~$H(z)$ data is limited to a redshift of 2.
Therefore, we limited ANN $H(z)$ to redshifts around 2 \cite{Mukherjee(2022)}.
Consequently, we exclusively selected GRB data with redshifts less than 2 for our analysis.

\section{Results and Discussion}\label{section3}

\begin{figure}
  \begin{center}
  \includegraphics[width=0.6\hsize]{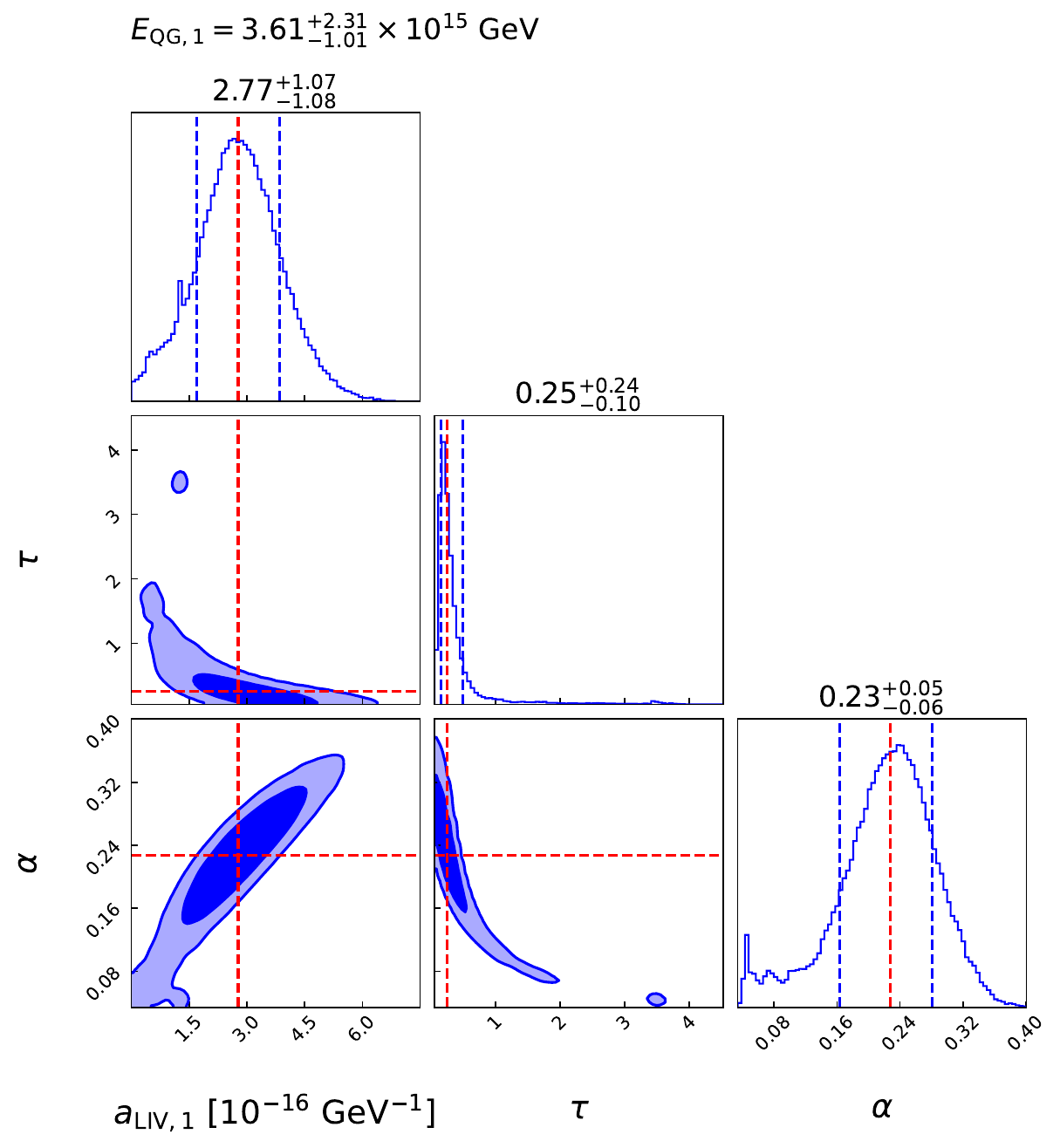}
  \end{center}
  \caption{The 1D probability distribution of each parameter and the 2D confidence contours for the parameters $a_{\mathrm{LIV},1}$, $\tau$, and $\alpha$ (the linear LIV case, i.e., n = 1). The blue dashed line delineates the $1\sigma$ confidence interval, whereas the red dashed line highlights the best-fit estimate. \label{example2}}
\end{figure}

\begin{figure}
  \begin{center}
  \includegraphics[width=0.6\hsize]{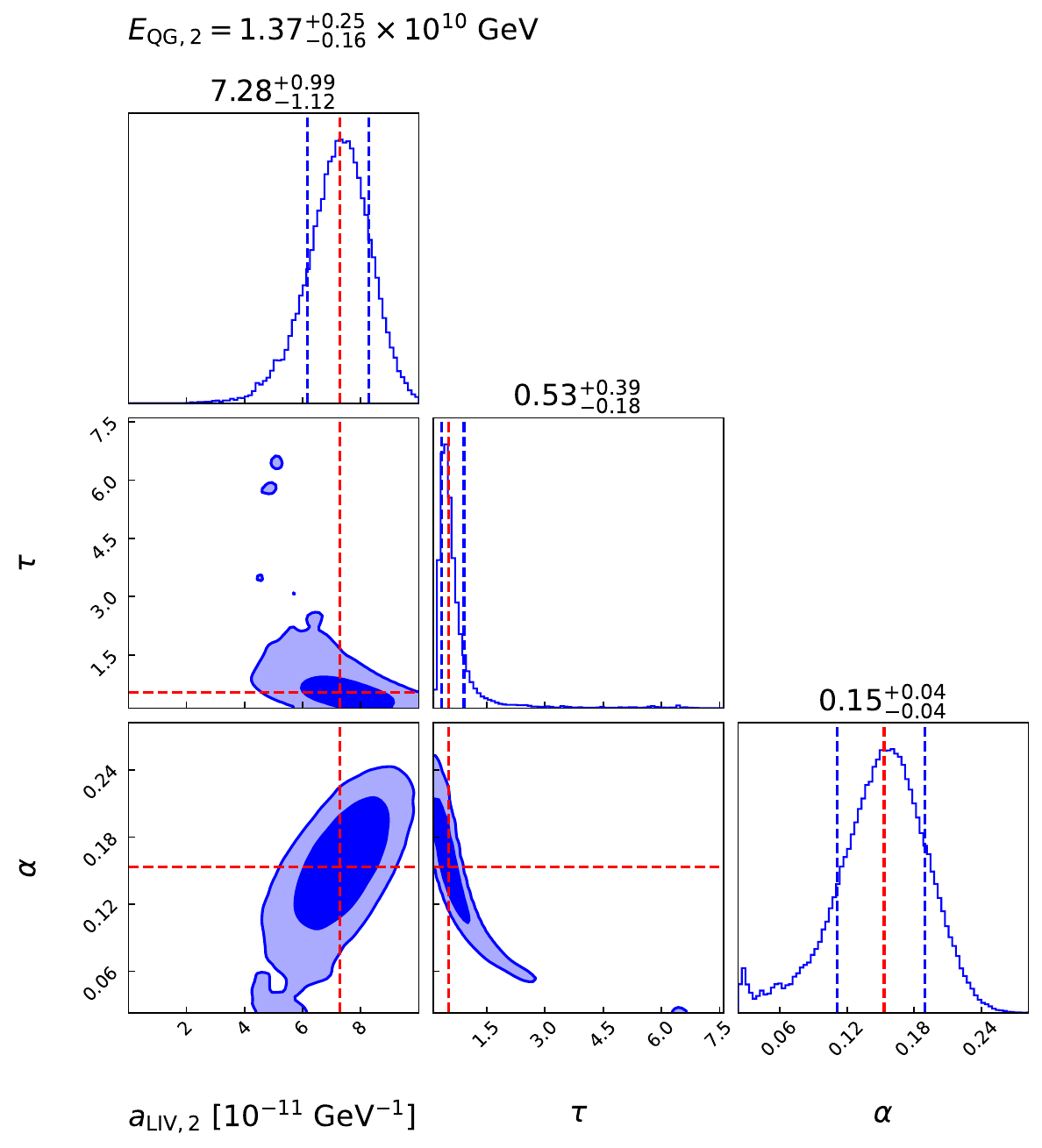}
  \end{center}
  \caption{The 1D probability distribution of each parameter and the 2D confidence contours for the parameters $a_{\mathrm{LIV},2}$, $\tau$, and $\alpha$ (the quadratic LIV case, i.e., n = 2).  \label{example3}}
\end{figure}

We have applied the ANN method to exclude the effects of the cosmological model.
Next, we employ the Monte Carlo Markov Chain method \cite{Lewis(2002)} to determine the LIV parameters ($a_{\mathrm{LIV},n}$, $\tau$, and $\alpha$) by minimizing the $\chi^2$ objective function:
\begin{equation}
\chi^{2}=\sum_{i=1}^{N_{\mathrm{GRB}} }\left [ \frac{\Delta t_{\mathrm{th}}\left ( a_{\mathrm{LIV},n}, \tau ,\alpha   \right ) -\Delta t_{\mathrm{obs}}  }{\sigma_{\mathrm{tot}} }  \right ]^{2},
\label{eq:12}
\end{equation}
where $\Delta t_{\mathrm{th}}$ represents the theoretical time delay, $\Delta t_{\mathrm{obs}}$ stands for the observed time delay, and $\sigma_{\mathrm{tot}}$ denotes the total uncertainty.
\begin{equation}
\sigma_{\mathrm{tot}}^{2} =\sigma_{t}^{2}+ \left( \frac{\partial f}{\partial E} \right) ^{2}\sigma_{E}^{2}+ \left( \frac{\partial f}{\partial E_{0}} \right)^{2}\sigma_{E_{0}}^{2},
\label{eq:13}
\end{equation}
where $f$ is Eq. \ref{eq:10}, $\sigma_{t}$ is the observed time delay uncertainty, and $\sigma_{E_{0}}$ and $\sigma_{E}$ represent the half-width of the lower and higher energy bands, respectively.
Table~\ref{tab:prior_values} lists the prior ranges of the parameters.
The posterior probability distribution for each parameter and the corresponding two-dimensional confidence contours are plotted in Figures \ref{example2} and \ref{example3}. 

\begin{table}[htbp]
\centering
\caption{Prior Ranges of Parameters}
\label{tab:prior_values}
\begin{tabular}{lcl}
\toprule
Parameter &  \multicolumn{2}{c}{Prior Range} \\
\midrule
$a_{\mathrm{LIV},1}$ & & $(0 \sim 30) \times 10^{-16} \ \mathrm{GeV}^{-1}$  \\
$a_{\mathrm{LIV},2}$ & & $(0\sim30) \times 10^{-11} \ \mathrm{GeV}^{-1}$ \\
$\tau$              & & $-20 \sim 20$ \\
$\alpha$            & & $-10 \sim 10$ \\
\bottomrule
\addlinespace
\end{tabular}
\end{table}

Given our limited understanding of GRB emission mechanisms and the inability to accurately characterize intrinsic time delays, we adopt a conservative interpretation of these results.
We begin our discussion with the linear LIV case (i.e., $n = 1$).
The constraints for $\tau$, $\alpha$, and $a_{\mathrm{LIV},1}$ are depicted in Figure \ref{example2}. The best-fit parameter values along with their corresponding $1\sigma$ uncertainties are $a_{\mathrm{LIV},1}=2.77_{-1.08}^{+1.07} \ (10^{-16} \ \mathrm{GeV}^{-1})$, $\tau=0.25_{-0.10}^{+0.24}$, and $\alpha=0.23_{-0.06}^{+0.05}$. Therefore, $E_{\mathrm{QG},1}=3.61_{-1.01}^{+2.31}\times10^{15} \ \mathrm{GeV}$, and the lower bound for LIV at the $1\sigma$ confidence level is set at $E_{\mathrm{QG},1}\ge 2.60\times 10^{15} \ \mathrm{GeV}$, a value that is four orders of magnitude beneath the Planck energy scale.
Although it is well below the Planck energy scale, it is similar to that of Xiao et al. \cite{Xiao(2022)}, who obtained $E_{\mathrm{QG},1} \ge 2.5 \times 10^{15} \ \mathrm{GeV}$ through an analysis of 46 SGRBs.
The best-fit values, $\tau \sim 0.25$ and $\alpha \sim 0.23$, both parameters lie within the $1\sigma$ range as positive values, indicating a positive intrinsic lag in our analysis.
Most GRB pulses are characterized by positive lags~\cite{Norris(2000),Yi(2006)}, our results align with the observational data.
From Eq.~\ref{eq:11}, the observed time delay is determined by the interplay between energy-dependent contributions from potential LIV effects and intrinsic emission mechanisms. These two components may exhibit different energy dependencies, leading to a varying net time delay across the spectrum. In some scenarios, this interplay could result in a change of the lag behavior with energy, such as a transition from negative to positive lags, depending on the relative strength of the contributions at low and high energies. While energy-dependent spectral variations of this kind have been reported in a few exceptional GRBs, such as GRB 160625B~\cite{Wei(2017)} and GRB 190114C~\cite{Du(2020)}, they are not commonly observed across the broader GRB population. In contrast to studies focusing on peculiar individual events, our analysis is based on a large sample of diverse GRBs. By combining multiple sources, we aim to place statistically robust constraints on the possible contribution of quantum gravity-induced LIV effects, without assuming specific emission models or strong spectral lag evolution in individual bursts.

In the quadratic LIV case (i.e., $n = 2$), the best-fit parameter values along with their corresponding $1\sigma$ uncertainties are  $a_{\mathrm{LIV},2}=7.28_{-1.12}^{+0.99} \ (10^{-11} \ \mathrm{GeV}^{-1})$, $\tau=0.53_{-0.18}^{+0.39}$, and $\alpha=0.15_{-0.04}^{+0.04}$.
The 1D probability distributions and 2D contours with $1\sigma$ and $2\sigma$ confidence levels for $\tau$, $\alpha$, and $a_{\mathrm{LIV},2}$ are also illustrated in Figure \ref{example3}.
The $1\sigma$ confidence-level lower boundary for LIV is $E_{\mathrm{QG},2} \geq 1.21 \times 10^{10} \mathrm{GeV}$, a value within the same order of magnitude as that obtained from the single high-energy GRB 190114C \cite{Acciari(2020)}.
The inferred values of $\tau$  and $\alpha$ are broadly consistent between the linear and quadratic LIV scenarios.
Additionally, a significant correlation is observed between the parameters $\tau$ and $\alpha$ in both scenarios.
Meanwhile, compared to previous analyses using the GP method \cite{Pan(2020)}, our ANN-based constraints show comparable precision.
It is not surprising that, to ensure a fully model-independence, we only utilize 32 CC $H(z)$, without taking into account the $H(z)$ obtained through the detection of radial BAO features.
Therefore, we hope that more $H(z)$ values can be measured to expand the redshift range of ANN $H(z)$ in the future.

These are the main reasons for our choice of the power-law model and the majority of time delay measurements of LGRBs. On one hand, it is not reasonable to assume the intrinsic lag as an unknown constant, especially when considering the fact that LGRBs exhibit long spectral lags.
Although some researchers exclusively use the constant model for a sample of SGRBs, which may be reasonable, primarily due to the negligible spectral lags of SGRBs \cite{Desai(2023)}.
However, considering the wide range of variations observed in GRB light curves \cite{Fishman(1995)}, assuming same time lags appears implausible and could potentially introduce systematic effects \cite{Desai(2023)}.
The power-law model was initially employed by Wei et al. \cite{Wei(2017)} to explain the transition in the spectral lag of GRBs, it is a function of energy.
Clearly, this model is more reasonable and provides a better description of intrinsic lags when compared to the constant model \cite{Wei(2017),Pan(2020)}.
On the other hand, the power-law model was deduced empirically by examining spectral lag characteristics in a dataset of 50 single-pulsed GRBs \cite{Shao(2017)}.
Please note that the dataset consists of only one short GRB, while the others are long GRBs. Lastly, a recent study by Desai et al. \cite{Desai(2023)} utilized this power-law model in 46 SGRBs to constrain LIV, but they did not achieve a favorable outcome.
The values of $\tau$ and $\alpha$ are centered around zero within a $1\sigma$ range, along with the inability to obtain a closed contour for $E_{\mathrm{QG}}$, possibly due to SGRBs' short spectral lag and different emission mechanism.

\section{Conclusions}\label{section4}

The search for possible LIV from astrophysical sources such as GRBs is essential for finding evidence of new theories of QG. We analyze 74 time delays from GRBs, including 37 measurements from GRB 160625B across various energy bands at redshift $z = 1.41$, and 37 additional GRBs with time delays spanning redshifts from $z = 0.117$ to $z = 1.99$. However, the effect of the underlying cosmological model is still understudied in the previous analysis. In this work, we adopt a novel approach by utilizing an ANN to reconstruct the cosmic expansion history. Specially, the newest observations of cosmic chronometers acting as the training set ensures a strict model-independent methodology. Our choice of training data encompasses 32 data points of the cosmic chronometer $H(z)$ (CC $H(z)$). This selection ensures that the resulting reconstruction of the cosmic expansion history remains independent of any specific theoretical model. Furthermore, we include the power-law model and discuss the searches for LIV in both linear and quadratic cases.

Using the power-law model and stacking of a large sample, we have established constraints for both linear and quadratic cases of LIV: $E_{\mathrm{QG},1} \ge 2.60 \times 10^{15} \ \mathrm{GeV}$ and $E_{\mathrm{QG},2} \ge 1.21 \times 10^{10} \ \mathrm{GeV}$. Notably, in the quadratic case, the value of $E_{\mathrm{QG},2}$ is within the same order of magnitude as that obtained from high-energy photons \cite{Acciari(2020)}. The parameters $\tau$ and $\alpha$ of the power-law model are within the $1\sigma$ range as positive values. Compared to using the time delays of a single special GRB, our findings present higher precision and improved constraints, as a larger sample of GRBs provides more robust limits on LIV. However, the precision of our ANN-based analysis is comparable to that achieved in previous studies using the GP method \cite{Pan(2020)}.We also elaborate on the rationale for adopting the power-law model and for selecting predominantly LGRBs in our sample. On the one hand, the power-law model is more physically motivated than the constant lag model, as it is an energy-dependent function capable of describing the transition in spectral lags \cite{Wei(2017),Du(2020)}. On the other hand, although some studies have applied this model to SGRBs, the resulting constraints on LIV were relatively weak \cite{Desai(2023)}.

Finally, the GRBs often have high redshifts and the reconstructed $H(z)$ is constrained within a range up to 2, we must exclude some data, which could introduce unforeseen effects on the results.
In the future, we hope to obtain more $H(z)$ values to extend the redshift range of ANN $H(z)$. More importantly, we have observed a significant correlation between $\tau$ and $\alpha$ in the power-law model, which may indicate a potential issue with this model.

\acknowledgments

We thank Shao-Qi Hou and Marek Biesiada for the helpful discussion on this paper. This work was supported by National Natural Science Foundation of China Grant Nos. 12105032, 12203009, 12021003, 11690023, and 11920101003; the Natural Science Foundation of Chongqing 18 No. cstc2021jcyj-msxmX0481; the Strategic Priority Research Program of the Chinese Academy of Sciences, Grant No. XDB23000000; the Central Guidance on Local Science and Technology Development Fund of Sichuan Province (24ZYZYTS0188);the Interdiscipline Research Funds of Beijing Normal University.

\bibliographystyle{JHEP}
\bibliography{biblio}

\providecommand{\href}[2]{#2}\begingroup\raggedright\begin{thebibliography}{10}

\bibitem{Mattingly(2005)}
D.~{Mattingly}, \emph{{Modern Tests of Lorentz Invariance}}, \href{https://doi.org/10.12942/lrr-2005-5}{\emph{Living Reviews in Relativity} {\bfseries 8} (2005) 5} [\href{https://arxiv.org/abs/gr-qc/0502097}{{\ttfamily gr-qc/0502097}}].

\bibitem{Amelino-Camelia(2013)}
G.~{Amelino-Camelia}, \emph{{Quantum-Spacetime Phenomenology}}, \href{https://doi.org/10.12942/lrr-2013-5}{\emph{Living Reviews in Relativity} {\bfseries 16} (2013) 5} [\href{https://arxiv.org/abs/0806.0339}{{\ttfamily 0806.0339}}].

\bibitem{Ellis(2008)}
J.~{Ellis}, N.E.~{Mavromatos}, D.V.~{Nanopoulos}, A.S.~{Sakharov} and E.K.G.~{Sarkisyan}, \emph{{Robust limits on Lorentz violation from gamma-ray bursts}}, \href{https://doi.org/10.1016/j.astropartphys.2006.04.001}{\emph{Astroparticle Physics} {\bfseries 25} (2006) 402} [\href{https://arxiv.org/abs/astro-ph/0510172}{{\ttfamily astro-ph/0510172}}].

\bibitem{Jacob(2008)}
U.~{Jacob} and T.~{Piran}, \emph{{Lorentz-violation-induced arrival delays of cosmological particles}}, \href{https://doi.org/10.1088/1475-7516/2008/01/031}{\emph{\jcap} {\bfseries 2008} (2008) 031} [\href{https://arxiv.org/abs/0712.2170}{{\ttfamily 0712.2170}}].

\bibitem{Desai(2023)}
S.~{Desai}, \emph{{Astrophysical and Cosmological Searches for Lorentz Invariance Violation}}, \href{https://doi.org/10.48550/arXiv.2303.10643}{\emph{arXiv:2303.10643} (2023) arXiv:2303.10643} [\href{https://arxiv.org/abs/2303.10643}{{\ttfamily 2303.10643}}].

\bibitem{Liao(2022)}
K.~{Liao}, M.~{Biesiada} and Z.-H.~{Zhu}, \emph{{Strongly Lensed Transient Sources: A Review}}, \href{https://doi.org/10.1088/0256-307X/39/11/119801}{\emph{Chinese Physics Letters} {\bfseries 39} (2022) 119801} [\href{https://arxiv.org/abs/2207.13489}{{\ttfamily 2207.13489}}].

\bibitem{Kouveliotou(1993)}
C.~{Kouveliotou}, C.A.~{Meegan}, G.J.~{Fishman}, N.P.~{Bhat}, M.S.~{Briggs}, T.M.~{Koshut} et~al., \emph{{Identification of Two Classes of Gamma-Ray Bursts}}, \href{https://doi.org/10.1086/186969}{\emph{Astrophysical Journal Letters} {\bfseries 413} (1993) L101}.

\bibitem{Nakar(2007)}
E.~{Nakar}, \emph{{Short-hard gamma-ray bursts}}, \href{https://doi.org/10.1016/j.physrep.2007.02.005}{\emph{Physical Review Letters} {\bfseries 442} (2007) 166} [\href{https://arxiv.org/abs/astro-ph/0701748}{{\ttfamily astro-ph/0701748}}].

\bibitem{Woosley(2006)}
S.E.~{Woosley} and J.S.~{Bloom}, \emph{{The Supernova Gamma-Ray Burst Connection}}, \href{https://doi.org/10.1146/annurev.astro.43.072103.150558}{\emph{Annual Review of Astronomy and Astrophysics} {\bfseries 44} (2006) 507} [\href{https://arxiv.org/abs/astro-ph/0609142}{{\ttfamily astro-ph/0609142}}].

\bibitem{Chen(2005)}
L.~{Chen}, Y.-Q.~{Lou}, M.~{Wu}, J.-L.~{Qu}, S.-M.~{Jia} and X.-J.~{Yang}, \emph{{Distribution of Spectral Lags in Gamma-Ray Bursts}}, \href{https://doi.org/10.1086/426774}{\emph{\apj} {\bfseries 619} (2005) 983} [\href{https://arxiv.org/abs/astro-ph/0410344}{{\ttfamily astro-ph/0410344}}].

\bibitem{Ukwatta(2012)}
T.N.~{Ukwatta}, K.S.~{Dhuga}, M.~{Stamatikos}, C.D.~{Dermer}, T.~{Sakamoto}, E.~{Sonbas} et~al., \emph{{The lag-luminosity relation in the GRB source frame: an investigation with Swift BAT bursts}}, \href{https://doi.org/10.1111/j.1365-2966.2011.19723.x}{\emph{\mnras} {\bfseries 419} (2012) 614} [\href{https://arxiv.org/abs/1109.0666}{{\ttfamily 1109.0666}}].

\bibitem{Bernardini(2015)}
M.G.~{Bernardini}, G.~{Ghirlanda}, S.~{Campana}, S.~{Covino}, R.~{Salvaterra}, J.L.~{Atteia} et~al., \emph{{Comparing the spectral lag of short and long gamma-ray bursts and its relation with the luminosity}}, \href{https://doi.org/10.1093/mnras/stu2153}{\emph{\mnras} {\bfseries 446} (2015) 1129} [\href{https://arxiv.org/abs/1410.5216}{{\ttfamily 1410.5216}}].

\bibitem{Xiao(2022)}
S.~{Xiao}, S.-L.~{Xiong}, Y.~{Wang}, S.-N.~{Zhang}, H.~{Gao}, Z.~{Zhang} et~al., \emph{{A Robust Estimation of Lorentz Invariance Violation and Intrinsic Spectral Lag of Short Gamma-Ray Bursts}}, \href{https://doi.org/10.3847/2041-8213/ac478a}{\emph{Astrophysical Journal Letters} {\bfseries 924} (2022) L29}.

\bibitem{Abdo(2009)A}
A.A.~{Abdo}, M.~{Ackermann}, M.~{Ajello}, K.~{Asano}, {Fermi LAT Collaboration} et~al., \emph{{A limit on the variation of the speed of light arising from quantum gravity effects}}, \href{https://doi.org/10.1038/nature08574}{\emph{\nat} {\bfseries 462} (2009) 331} [\href{https://arxiv.org/abs/0908.1832}{{\ttfamily 0908.1832}}].

\bibitem{Abdo(2009)B}
A.A.~{Abdo}, M.~{Ackermann}, M.~{Arimoto}, {Fermi LAT Collaboration}, {Fermi GBM Collaboration} et~al., \emph{{Fermi Observations of High-Energy Gamma-Ray Emission from GRB 080916C}}, \href{https://doi.org/10.1126/science.1169101}{\emph{Science} {\bfseries 323} (2009) 1688}.

\bibitem{Acciari(2020)}
V.A.~{Acciari}, S.~{Ansoldi}, L.A.~{Antonelli}, {MAGIC Collaboration} et~al., \emph{{Bounds on Lorentz Invariance Violation from MAGIC Observation of GRB 190114C}}, \href{https://doi.org/10.1103/PhysRevLett.125.021301}{\emph{\prl} {\bfseries 125} (2020) 021301} [\href{https://arxiv.org/abs/2001.09728}{{\ttfamily 2001.09728}}].

\bibitem{Wei(2017)}
J.-J.~{Wei}, B.-B.~{Zhang}, L.~{Shao}, X.-F.~{Wu} and P.~{M{\'e}sz{\'a}ros}, \emph{{A New Test of Lorentz Invariance Violation: The Spectral Lag Transition of GRB 160625B}}, \href{https://doi.org/10.3847/2041-8213/834/2/L13}{\emph{Astrophysical Journal Letters} {\bfseries 834} (2017) L13} [\href{https://arxiv.org/abs/1612.09425}{{\ttfamily 1612.09425}}].

\bibitem{Agrawal(2021)}
R.~{Agrawal}, H.~{Singirikonda} and S.~{Desai}, \emph{{Search for Lorentz Invariance Violation from stacked Gamma-Ray Burst spectral lag data}}, \href{https://doi.org/10.1088/1475-7516/2021/05/029}{\emph{\jcap} {\bfseries 2021} (2021) 029} [\href{https://arxiv.org/abs/2102.11248}{{\ttfamily 2102.11248}}].

\bibitem{Pan(2020)}
Y.~{Pan}, J.~{Qi}, S.~{Cao}, T.~{Liu}, Y.~{Liu}, S.~{Geng} et~al., \emph{{Model-independent Constraints on Lorentz Invariance Violation: Implication from Updated Gamma-Ray Burst Observations}}, \href{https://doi.org/10.3847/1538-4357/ab6ef5}{\emph{\apj} {\bfseries 890} (2020) 169} [\href{https://arxiv.org/abs/2001.08451}{{\ttfamily 2001.08451}}].

\bibitem{2025arXiv250314739D}
{DESI Collaboration}, M.~{Abdul-Karim}, J.~{Aguilar}, S.~{Ahlen}, C.~{Allende Prieto}, O.~{Alves} et~al., \emph{{DESI DR2 Results I: Baryon Acoustic Oscillations from the Lyman Alpha Forest}}, \href{https://doi.org/10.48550/arXiv.2503.14739}{\emph{arXiv e-prints} (2025) arXiv:2503.14739} [\href{https://arxiv.org/abs/2503.14739}{{\ttfamily 2503.14739}}].

\bibitem{2025arXiv250314738D}
{DESI Collaboration}, M.~{Abdul-Karim}, J.~{Aguilar}, S.~{Ahlen}, S.~{Alam}, L.~{Allen} et~al., \emph{{DESI DR2 Results II: Measurements of Baryon Acoustic Oscillations and Cosmological Constraints}}, \href{https://doi.org/10.48550/arXiv.2503.14738}{\emph{arXiv e-prints} (2025) arXiv:2503.14738} [\href{https://arxiv.org/abs/2503.14738}{{\ttfamily 2503.14738}}].

\bibitem{2025arXiv250314743L}
K.~{Lodha}, R.~{Calderon}, W.L.~{Matthewson}, A.~{Shafieloo}, M.~{Ishak}, J.~{Pan} et~al., \emph{{Extended Dark Energy analysis using DESI DR2 BAO measurements}}, \href{https://doi.org/10.48550/arXiv.2503.14743}{\emph{arXiv e-prints} (2025) arXiv:2503.14743} [\href{https://arxiv.org/abs/2503.14743}{{\ttfamily 2503.14743}}].

\bibitem{Staicova(2023)}
D.~{Staicova}, \emph{{Impact of cosmology on Lorentz Invariance Violation constraints from GRB time-delays}}, \href{https://doi.org/10.1088/1361-6382/acf270}{\emph{Classical and Quantum Gravity} {\bfseries 40} (2023) 195012} [\href{https://arxiv.org/abs/2305.06504}{{\ttfamily 2305.06504}}].

\bibitem{2018ApJ...856....3Y}
H.~{Yu}, B.~{Ratra} and F.-Y.~{Wang}, \emph{{Hubble Parameter and Baryon Acoustic Oscillation Measurement Constraints on the Hubble Constant, the Deviation from the Spatially Flat {\ensuremath{\Lambda}}CDM Model, the Deceleration-Acceleration Transition Redshift, and Spatial Curvature}}, \href{https://doi.org/10.3847/1538-4357/aab0a2}{\emph{\apj} {\bfseries 856} (2018) 3} [\href{https://arxiv.org/abs/1711.03437}{{\ttfamily 1711.03437}}].

\bibitem{2019ApJ...886...94L}
T.~{Liu}, S.~{Cao}, J.~{Zhang}, S.~{Geng}, Y.~{Liu}, X.~{Ji} et~al., \emph{{Implications from Simulated Strong Gravitational Lensing Systems: Constraining Cosmological Parameters Using Gaussian Processes}}, \href{https://doi.org/10.3847/1538-4357/ab4bc3}{\emph{\apj} {\bfseries 886} (2019) 94} [\href{https://arxiv.org/abs/1910.02592}{{\ttfamily 1910.02592}}].

\bibitem{2023MNRAS.523.3406F}
A.~{Favale}, A.~{G{\'o}mez-Valent} and M.~{Migliaccio}, \emph{{Cosmic chronometers to calibrate the ladders and measure the curvature of the Universe. A model-independent study}}, \href{https://doi.org/10.1093/mnras/stad1621}{\emph{\mnras} {\bfseries 523} (2023) 3406} [\href{https://arxiv.org/abs/2301.09591}{{\ttfamily 2301.09591}}].

\bibitem{Mukherjee(2022)}
P.~{Mukherjee}, J.L.~{Said} and J.~{Mifsud}, \emph{{Neural network reconstruction of H'(z) and its application in teleparallel gravity}}, \href{https://doi.org/10.1088/1475-7516/2022/12/029}{\emph{\jcap} {\bfseries 2022} (2022) 029} [\href{https://arxiv.org/abs/2209.01113}{{\ttfamily 2209.01113}}].

\bibitem{Zhou(2019)}
H.~{Zhou} and Z.~{Li}, \emph{{Testing the fidelity of Gaussian processes for cosmography}}, \href{https://doi.org/10.1088/1674-1137/43/3/035103}{\emph{Chinese Physics C} {\bfseries 43} (2019) 035103}.

\bibitem{Wang(2020)}
G.-J.~{Wang}, X.-J.~{Ma}, S.-Y.~{Li} and J.-Q.~{Xia}, \emph{{Reconstructing Functions and Estimating Parameters with Artificial Neural Networks: A Test with a Hubble Parameter and SNe Ia}}, \href{https://doi.org/10.3847/1538-4365/ab620b}{\emph{\apjs} {\bfseries 246} (2020) 13} [\href{https://arxiv.org/abs/1910.03636}{{\ttfamily 1910.03636}}].

\bibitem{Go'mez-Vargas(2023)}
I.~{G{\'o}mez-Vargas}, R.~{Medel-Esquivel}, R.~{Garc{\'\i}a-Salcedo} and J.A.~{V{\'a}zquez}, \emph{{Neural network reconstructions for the Hubble parameter, growth rate and distance modulus}}, \href{https://doi.org/10.1140/epjc/s10052-023-11435-9}{\emph{European Physical Journal C} {\bfseries 83} (2023) 304}.

\bibitem{Qi(2023)}
J.-Z.~{Qi}, P.~{Meng}, J.-F.~{Zhang} and X.~{Zhang}, \emph{{Model-independent measurement of cosmic curvature with the latest H (z ) and SNe Ia data: A comprehensive investigation}}, \href{https://doi.org/10.1103/PhysRevD.108.063522}{\emph{\prd} {\bfseries 108} (2023) 063522} [\href{https://arxiv.org/abs/2302.08889}{{\ttfamily 2302.08889}}].

\bibitem{1998Natur.395Q.525A}
G.~{Amelino-Camelia}, J.~{Ellis}, N.E.~{Mavromatos}, D.V.~{Nanopoulos} and S.~{Sarkar}, \emph{{Tests of quantum gravity from observations of {\ensuremath{\gamma}}-ray bursts}}, \href{https://doi.org/10.1038/26793}{\emph{\nat} {\bfseries 395} (1998) 525}.

\bibitem{2008ApJ...689L...1K}
V.A.~{Kosteleck{\'y}} and M.~{Mewes}, \emph{{Astrophysical Tests of Lorentz and CPT Violation with Photons}}, \href{https://doi.org/10.1086/595815}{\emph{\apjl} {\bfseries 689} (2008) L1} [\href{https://arxiv.org/abs/0809.2846}{{\ttfamily 0809.2846}}].

\bibitem{2022PrPNP.12503948A}
A.~{Addazi}, J.~{Alvarez-Muniz}, R.~{Alves Batista}, G.~{Amelino-Camelia}, V.~{Antonelli}, M.~{Arzano} et~al., \emph{{Quantum gravity phenomenology at the dawn of the multi-messenger era-A review}}, \href{https://doi.org/10.1016/j.ppnp.2022.103948}{\emph{Progress in Particle and Nuclear Physics} {\bfseries 125} (2022) 103948} [\href{https://arxiv.org/abs/2111.05659}{{\ttfamily 2111.05659}}].

\bibitem{2013PhRvD..87l2001V}
V.~{Vasileiou}, A.~{Jacholkowska}, F.~{Piron}, J.~{Bolmont}, C.~{Couturier}, J.~{Granot} et~al., \emph{{Constraints on Lorentz invariance violation from Fermi-Large Area Telescope observations of gamma-ray bursts}}, \href{https://doi.org/10.1103/PhysRevD.87.122001}{\emph{\prd} {\bfseries 87} (2013) 122001} [\href{https://arxiv.org/abs/1305.3463}{{\ttfamily 1305.3463}}].

\bibitem{1995AnPhy.243...90L}
J.~{Lukierski}, H.~{Ruegg} and W.J.~{Zakrzewski}, \emph{{Classical and Quantum Mechanics of Free {\ensuremath{\kappa}}-Relativistic Systems}}, \href{https://doi.org/10.1006/aphy.1995.1092}{\emph{Annals of Physics} {\bfseries 243} (1995) 90} [\href{https://arxiv.org/abs/hep-th/9312153}{{\ttfamily hep-th/9312153}}].

\bibitem{1997IJMPA..12..607A}
G.~{Amelino-Camelia}, J.~{Ellis}, N.E.~{Mavromatos} and D.V.~{Nanopoulos}, \emph{{Distance Measurement and Wave Dispersion in a Liouville-String Approach to Quantum Gravity}}, \href{https://doi.org/10.1142/S0217751X97000566}{\emph{International Journal of Modern Physics A} {\bfseries 12} (1997) 607} [\href{https://arxiv.org/abs/hep-th/9605211}{{\ttfamily hep-th/9605211}}].

\bibitem{1998AJ....116.1009R}
A.G.~{Riess}, A.V.~{Filippenko}, P.~{Challis}, A.~{Clocchiatti}, A.~{Diercks}, P.M.~{Garnavich} et~al., \emph{{Observational Evidence from Supernovae for an Accelerating Universe and a Cosmological Constant}}, \href{https://doi.org/10.1086/300499}{\emph{\aj} {\bfseries 116} (1998) 1009} [\href{https://arxiv.org/abs/astro-ph/9805201}{{\ttfamily astro-ph/9805201}}].

\bibitem{Zhang(2015)}
S.~{Zhang} and B.-Q.~{Ma}, \emph{{Lorentz violation from gamma-ray bursts}}, \href{https://doi.org/10.1016/j.astropartphys.2014.04.008}{\emph{Astroparticle Physics} {\bfseries 61} (2015) 108} [\href{https://arxiv.org/abs/1406.4568}{{\ttfamily 1406.4568}}].

\bibitem{Gao(2015)}
H.~{Gao}, X.-F.~{Wu} and P.~{M{\'e}sz{\'a}ros}, \emph{{Cosmic Transients Test Einstein's Equivalence Principle out to GeV Energies}}, \href{https://doi.org/10.1088/0004-637X/810/2/121}{\emph{\apj} {\bfseries 810} (2015) 121} [\href{https://arxiv.org/abs/1509.00150}{{\ttfamily 1509.00150}}].

\bibitem{Wei(2015)}
J.-J.~{Wei}, H.~{Gao}, X.-F.~{Wu} and P.~{M{\'e}sz{\'a}ros}, \emph{{Testing Einstein's Equivalence Principle With Fast Radio Bursts}}, \href{https://doi.org/10.1103/PhysRevLett.115.261101}{\emph{\prl} {\bfseries 115} (2015) 261101} [\href{https://arxiv.org/abs/1512.07670}{{\ttfamily 1512.07670}}].

\bibitem{2024ApJ...975L..29V}
M.K.~{Vyas}, A.~{Pe'er} and S.~{Iyyani}, \emph{{Unified Theory of Negative and Positive Spectral Lags in the Gamma-Ray Burst Prompt Phase due to Shear Comptonization from a Structured Jet}}, \href{https://doi.org/10.3847/2041-8213/ad887c}{\emph{\apjl} {\bfseries 975} (2024) L29} [\href{https://arxiv.org/abs/2410.04446}{{\ttfamily 2410.04446}}].

\bibitem{zhang(2017)}
B.B.~{Zhang}, B.~{Zhang}, A.J.~{Castro-Tirado} et~al., \emph{{Transition from fireball to Poynting-flux-dominated outflow in the three-episode GRB 160625B}}, \href{https://doi.org/10.1038/s41550-017-0309-8}{\emph{Nature Astronomy} {\bfseries 2} (2018) 69} [\href{https://arxiv.org/abs/1612.03089}{{\ttfamily 1612.03089}}].

\bibitem{Evans(2009)}
P.A.~{Evans}, A.P.~{Beardmore}, K.L.~{Page} et~al., \emph{{Methods and results of an automatic analysis of a complete sample of Swift-XRT observations of GRBs}}, \href{https://doi.org/10.1111/j.1365-2966.2009.14913.x}{\emph{\mnras} {\bfseries 397} (2009) 1177} [\href{https://arxiv.org/abs/0812.3662}{{\ttfamily 0812.3662}}].

\bibitem{Kienlin(2020)}
A.~{von Kienlin}, C.A.~{Meegan}, W.S.~{Paciesas} et~al., \emph{{The Fourth Fermi-GBM Gamma-Ray Burst Catalog: A Decade of Data}}, \href{https://doi.org/10.3847/1538-4357/ab7a18}{\emph{\apj} {\bfseries 893} (2020) 46} [\href{https://arxiv.org/abs/2002.11460}{{\ttfamily 2002.11460}}].

\bibitem{HORNIK1989359}
K.~Hornik, M.~Stinchcombe and H.~White, \emph{Multilayer feedforward networks are universal approximators}, \href{https://doi.org/https://doi.org/10.1016/0893-6080(89)90020-8}{\emph{Neural Networks} {\bfseries 2} (1989) 359}.

\bibitem{Jimenez(2002)}
R.~{Jimenez} and A.~{Loeb}, \emph{{Constraining Cosmological Parameters Based on Relative Galaxy Ages}}, \href{https://doi.org/10.1086/340549}{\emph{\apj} {\bfseries 573} (2002) 37} [\href{https://arxiv.org/abs/astro-ph/0106145}{{\ttfamily astro-ph/0106145}}].

\bibitem{Blake(2012)}
E.~{Gazta{\~n}aga}, A.~{Cabr{\'e}} and L.~{Hui}, \emph{{Clustering of luminous red galaxies - IV. Baryon acoustic peak in the line-of-sight direction and a direct measurement of H(z)}}, \href{https://doi.org/10.1111/j.1365-2966.2009.15405.x}{\emph{\mnras} {\bfseries 399} (2009) 1663} [\href{https://arxiv.org/abs/0807.3551}{{\ttfamily 0807.3551}}].

\bibitem{Samushia(2013)}
L.~{Samushia}, B.A.~{Reid}, M.~{White} et~al., \emph{{The clustering of galaxies in the SDSS-III DR9 Baryon Oscillation Spectroscopic Survey: testing deviations from {\ensuremath{\Lambda}} and general relativity using anisotropic clustering of galaxies}}, \href{https://doi.org/10.1093/mnras/sts443}{\emph{\mnras} {\bfseries 429} (2013) 1514} [\href{https://arxiv.org/abs/1206.5309}{{\ttfamily 1206.5309}}].

\bibitem{Zhang(2014)}
C.~{Zhang}, H.~{Zhang}, S.~{Yuan}, S.~{Liu}, T.-J.~{Zhang} and Y.-C.~{Sun}, \emph{{Four new observational H(z) data from luminous red galaxies in the Sloan Digital Sky Survey data release seven}}, \href{https://doi.org/10.1088/1674-4527/14/10/002}{\emph{Research in Astronomy and Astrophysics} {\bfseries 14} (2014) 1221} [\href{https://arxiv.org/abs/1207.4541}{{\ttfamily 1207.4541}}].

\bibitem{Stern(2010)}
D.~{Stern}, R.~{Jimenez}, L.~{Verde}, M.~{Kamionkowski} and S.A.~{Stanford}, \emph{{Cosmic chronometers: constraining the equation of state of dark energy. I: H(z) measurements}}, \href{https://doi.org/10.1088/1475-7516/2010/02/008}{\emph{\jcap} {\bfseries 2010} (2010) 008} [\href{https://arxiv.org/abs/0907.3149}{{\ttfamily 0907.3149}}].

\bibitem{Moresco(2012)}
M.~{Moresco}, A.~{Cimatti}, R.~{Jimenez} et~al., \emph{{Improved constraints on the expansion rate of the Universe up to z \raisebox{-0.5ex}\textasciitilde 1.1 from the spectroscopic evolution of cosmic chronometers}}, \href{https://doi.org/10.1088/1475-7516/2012/08/006}{\emph{\jcap} {\bfseries 2012} (2012) 006} [\href{https://arxiv.org/abs/1201.3609}{{\ttfamily 1201.3609}}].

\bibitem{Moresco(2016)}
M.~{Moresco}, L.~{Pozzetti}, A.~{Cimatti}, R.~{Jimenez}, C.~{Maraston}, L.~{Verde} et~al., \emph{{A 6\% measurement of the Hubble parameter at z\raisebox{-0.5ex}\textasciitilde0.45: direct evidence of the epoch of cosmic re-acceleration}}, \href{https://doi.org/10.1088/1475-7516/2016/05/014}{\emph{\jcap} {\bfseries 2016} (2016) 014} [\href{https://arxiv.org/abs/1601.01701}{{\ttfamily 1601.01701}}].

\bibitem{Ratsimbazafy(2017)}
A.L.~{Ratsimbazafy}, S.I.~{Loubser}, S.M.~{Crawford}, C.M.~{Cress}, B.A.~{Bassett}, R.C.~{Nichol} et~al., \emph{{Age-dating luminous red galaxies observed with the Southern African Large Telescope}}, \href{https://doi.org/10.1093/mnras/stx301}{\emph{\mnras} {\bfseries 467} (2017) 3239} [\href{https://arxiv.org/abs/1702.00418}{{\ttfamily 1702.00418}}].

\bibitem{Aghanim(2021)}
{Planck Collaboration}, N.~{Aghanim}, Y.~{Akrami}, M.~{Ashdown} et~al., \emph{{Planck 2018 results. VI. Cosmological parameters}}, \href{https://doi.org/10.1051/0004-6361/201833910}{\emph{\aap} {\bfseries 641} (2020) A6} [\href{https://arxiv.org/abs/1807.06209}{{\ttfamily 1807.06209}}].

\bibitem{Lewis(2002)}
A.~{Lewis} and S.~{Bridle}, \emph{{Cosmological parameters from CMB and other data: A Monte Carlo approach}}, \href{https://doi.org/10.1103/PhysRevD.66.103511}{\emph{\prd} {\bfseries 66} (2002) 103511} [\href{https://arxiv.org/abs/astro-ph/0205436}{{\ttfamily astro-ph/0205436}}].

\bibitem{Norris(2000)}
J.P.~{Norris}, G.F.~{Marani} and J.T.~{Bonnell}, \emph{{Connection between Energy-dependent Lags and Peak Luminosity in Gamma-Ray Bursts}}, \href{https://doi.org/10.1086/308725}{\emph{\apj} {\bfseries 534} (2000) 248} [\href{https://arxiv.org/abs/astro-ph/9903233}{{\ttfamily astro-ph/9903233}}].

\bibitem{Yi(2006)}
T.~{Yi}, E.~{Liang}, Y.~{Qin} and R.~{Lu}, \emph{{On the spectral lags of the short gamma-ray bursts}}, \href{https://doi.org/10.1111/j.1365-2966.2006.10083.x}{\emph{\mnras} {\bfseries 367} (2006) 1751} [\href{https://arxiv.org/abs/astro-ph/0512270}{{\ttfamily astro-ph/0512270}}].

\bibitem{Du(2020)}
S.-S.~{Du}, L.~{Lan}, J.-J.~{Wei}, Z.-M.~{Zhou}, H.~{Gao}, L.-Y.~{Jiang} et~al., \emph{{Lorentz Invariance Violation Limits from the Spectral-lag Transition of GRB 190114C}}, \href{https://doi.org/10.3847/1538-4357/abc624}{\emph{\apj} {\bfseries 906} (2021) 8} [\href{https://arxiv.org/abs/2010.16029}{{\ttfamily 2010.16029}}].

\bibitem{Fishman(1995)}
G.J.~{Fishman}, \emph{{Gamma-Ray Bursts: an Overview}}, \href{https://doi.org/10.1086/133672}{\emph{Publications of the Astronomical Society of the Pacific} {\bfseries 107} (1995) 1145}.

\bibitem{Shao(2017)}
L.~{Shao}, B.-B.~{Zhang}, F.-R.~{Wang} et~al., \emph{{A New Measurement of the Spectral Lag of Gamma-Ray Bursts and its Implications for Spectral Evolution Behaviors}}, \href{https://doi.org/10.3847/1538-4357/aa7d01}{\emph{\apj} {\bfseries 844} (2017) 126} [\href{https://arxiv.org/abs/1610.07191}{{\ttfamily 1610.07191}}].

\end{thebibliography}\endgroup

\end{document}